\documentclass[a4paper,10pt]{cpc-hepnp}

\usepackage{multicol}
\usepackage{lineno}
\usepackage{graphicx}
\usepackage{booktabs}
\usepackage{amssymb,bm,mathrsfs,bbm,amscd}
\usepackage[tbtags]{amsmath}
\usepackage{lastpage}

\begin{document}

\fancyhead[c]{\small Preprint submitted to Chinese Physics C
} \fancyfoot[C]{\small Page-\thepage}


\title{Study on transient beam loading compensation for China ADS proton linac injector II\thanks{Supported by National Natural Science
Foundation of China (91426303),(11525523) }}

\author{%
      GAO Zheng$^{1,2;1)}$\email{gaozheng@impcas.ac.cn}%
\quad HE Yuan$^{1}$%
\quad WANG Xian-Wu$^{1}$
\quad CHANG Wei$^{1}$ \\
\quad ZHANG Rui-Feng$^{1}$
\quad ZHU Zheng-Long$^{1}$
\quad ZHANG Sheng-Hu$^{1}$ \\
\quad CHEN Qi$^{1,2}$
\quad Tom Powers$^{3}$
}

\maketitle

\address{%
$^1$ Institute of Modern Physics, Chinese Academy of Sciences, Lanzhou 730000, China\\
$^2$ Graduate University of Chinese Academy of Sciences, Beijing 100049, China\\
$^3$ Thomas Jefferson National Accelerator Facility, Newport News, VA 23606, USA\\
}

\begin{abstract}
Significant transient beam loading effects were observed during beam commissioning tests  of prototype II of the injector for the Accelerator Driven Sub-critical (ADS) system, which took place at the Institute of Modern Physics, Chinese Academy of Sciences, between October and December 2014. During these tests experiments were performed with CW operation of the cavities with pulsed beam current, and the system was configured to make use of a prototype digital low level radio frequency (LLRF) controller.  The system was originally operated in pulsed mode with a simple PID  feedback control algorithm, which was not able to maintain the desired gradient regulation during pulsed 10 mA beam operations. A unique simple transient beam loading compensation method which made use of a combination of PI feedback and feedforward control algorithm was implemented in order to significantly reduce the beam induced transient effect in the cavity gradients. The superconducting cavity field variation was reduced to less than 1.7\% after turning on this control algorithm. The design and experimental results of this system are presented in this paper.
\end{abstract}

\begin{keyword}
beam loading, LLRF, feedforward, cavity, FPGA
\end{keyword}

\begin{pacs}
9.20.Ej, 29.27.-a
\end{pacs}




\section{Introduction}
The China ADS project proposes to build an accelerator driven sub-critical system, aiming to transmute long-lived radionuclides to short-lived radionuclides. The accelerator will accelerate a proton beam to about 1 GeV to produce high flux neutrons for transmutation of radioactive nuclear waste~\cite{lab1,lab2}.

The accelerator for ADS is a superconducting (SC) proton linac operated in continuous-wave mode. Injector II is designed to accelerate the proton beam to 10 MeV with beam current up to 10 mA. This accelerator system include a proton ECR ion source, a low energy beam transport (LEBT) line, a 2.1 MeV room-temperature radio frequency quadrupole (RFQ), a medium energy beam transport (MEBT), and a superconducting radio frequency (SRF) linac system~\cite{lab3,lab4}. The beam commissioning began with a pulsed low current proton beam, then the beam current and pulse length were increased to achieve continuous high current beam acceleration.  In both cases the RF system was operated in CW mode. The superconducting linac prototype, which contained one test cryomodule and was capable of accelerating the proton beam by 2.5 MeV, was installed on the beam line.

The interaction of the beam with the cavity fields in accelerating mode are referred to as beam loading~\cite{lab5}. The proton beam of ADS Injector II was pulsed during beam commissioning, which caused transient beam loading that initially resulted in a significant transient effect in the accelerating gradients.  This degraded the beam quality and, at times, caused beam loss. The beam test results indicated that the beam loading effect should be compensated.

The LLRF control system is the key component to regulate RF field in the cavity. It is used as a closed loop controller which maintains cavity gradient and phase stability when operating the cavity with beam.  Digital and analog control systems have been developed at any number of institutions for control of RF cavities \cite{lab6,lab7,lab8,lab9,lab10} for both CW and pulsed operations in charged particle accelerators.

The adaptive feedforward control algorithm is a useful method to compensate for beam loading effects. One recent work which describes the theory and simulation is presented in reference \cite{lab11}. The algorithm is also deployed in some LLRF control systems~\cite{lab12,lab13}, but its drawback is that after turning on the algorithm to recover the cavity voltage, the learning cycles introduce more delays, which means the cavity voltage drops during the cycles at the beginning.

\section{The LLRF control system}
A block diagram of the cavity RF system used for this work is shown in Fig.~\ref{fig1}.  Not shown is an RF master oscillator and distribution system which is used to provide a frequency and phase reference for the systems. The LLRF control system, which is mounted in a single chassis, was developed by the Institute of Modern Physics, Chinese Academy of Sciences (IMPCAS)~\cite{lab14}.  The digital feedback functions of this LLRF control system are based on the traditional amplitude and phase control.  The data is acquired and initially processed using synchronous in-phase and quadrature demodulation techniques~\cite{lab15}. The digital signal processing is implemented using an Altera Stratix III EP3SL150F1152 field programmable gate array (FPGA).

\begin{center}
\includegraphics[width=\columnwidth]{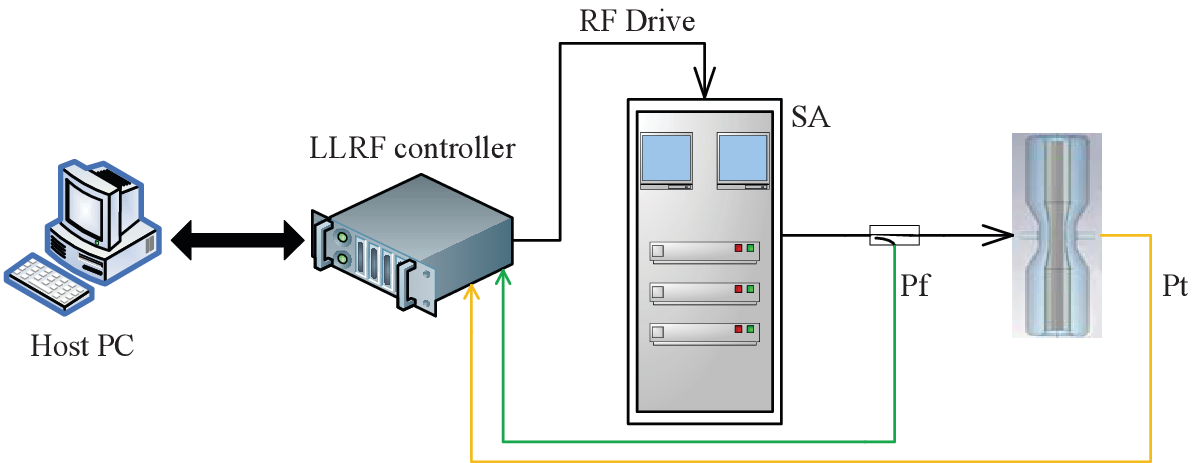}
\figcaption{\label{fig1} Block diagram of the RF control system.}
\end{center}

\section{Compensation algorithms}

\subsection{Feedback control algorithm}
The RF feedback control algorithm is realized as a positional Proportional-Integral (PI) controller. The output data of the PI controller is used to adjust the phase word of the NCO and modulate the amplitude of the NCO. The block diagram of the digital algorithm is shown in Fig.~\ref{fig2}. The input-output relationship of the PI controller can be expressed as Eq.(\ref{eq:1}).

\begin{equation}
\label{eq:1}
u(n) = {K_P}(e(n) + {K_I}\sum\limits_{i = 0}^n {e(i)} ),
\end{equation}
where $u(n)$ is the controller output, $K_P$ is proportional gain, $K_I$ is integral gain, and $e(n)$ is the error of measured amplitude or phase value from the setpoint.

\begin{center}
\includegraphics[width=0.8\columnwidth]{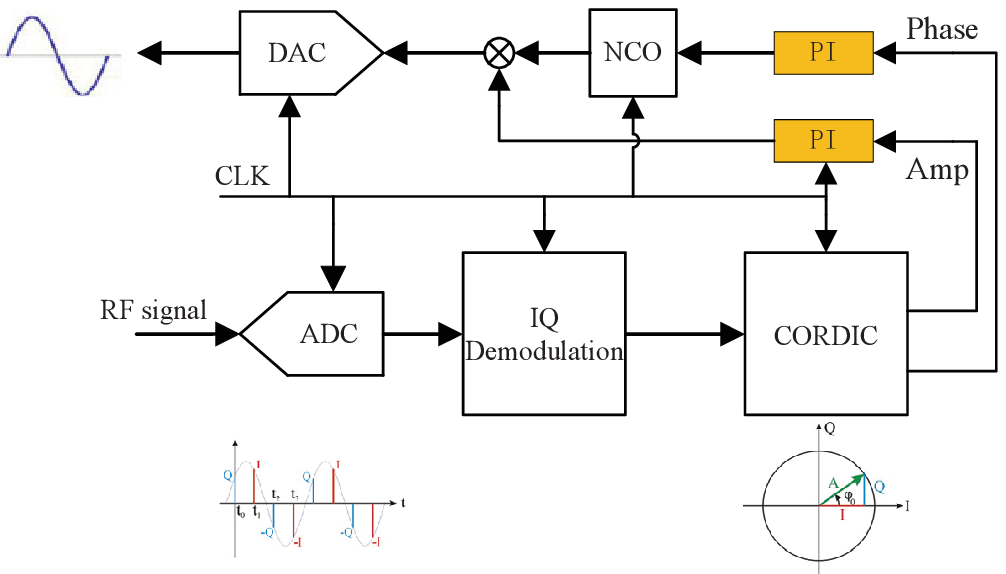}
\figcaption{\label{fig2} Block diagram of the digital algorithm.}
\end{center}

\subsection{Feedforward control algorithm}
The pulsed beam causes repetitive perturbation in the RF feedback control loop. Often  such a perturbation can be sufficiently compensated for using feedback. However in this case, because of the heavy beam loading and the bandwidth of the cavity-coupler system, the repetitive error cannot be eliminated by conventional closed-loop feedback techniques. In this case the PI controller barely recovers the cavity voltage by the end of the 1 ms beam pulse. The feedforward (FF) control techniques provide an effective way to correct repetitive errors.

The block diagram of the RF feedforward control algrorithm used in the LLRF system is shown in Fig.~\ref{fig3}. When enabled, the FF controller produces an array of correction values which is summed with the output of the PI controller during the beam pulse. This summed signal is used to drive the RF amplifier. The output of the FF control algorithm was initialized by setting the FF value from the control interface. The FF value was stored in the FPGA register, and the FF value can be adjusted according to the beam current. It also means that the output of the FF controller is adjusted manually by the operator.

The output of the feedback and feedforward controller can be simplified by the following Eq.(\ref{eq:2}).

\begin{equation}
\label{eq:2}
{u_f}(n) = {K_P}(e(n) + {K_I}\sum\limits_{i = 0}^n {e(i)} ) + f(n),
\end{equation}
where $f(n)$ is the output correction values of FF controller which is triggered on the rising edge of the beam pulse.

\begin{center}
\includegraphics[width=0.6\columnwidth]{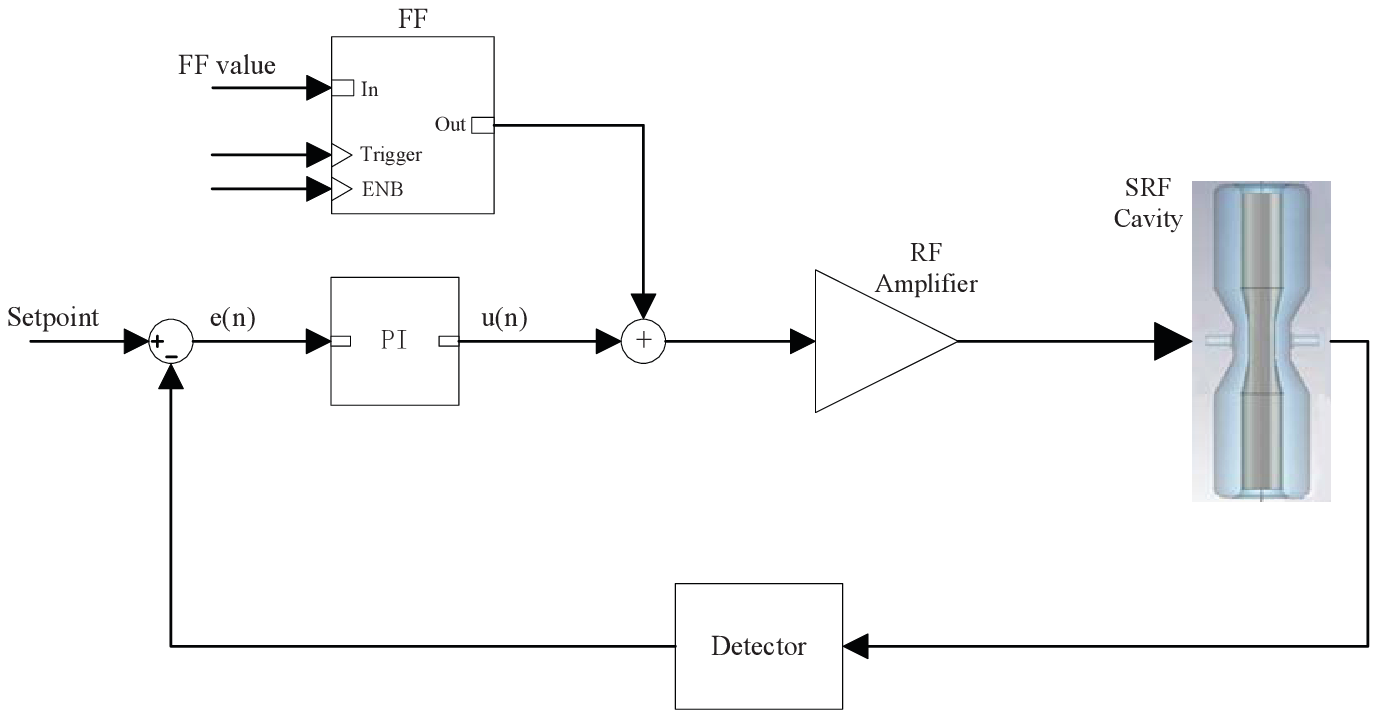}
\figcaption{\label{fig3} Block diagram of the IMPCAS RF feedforward control system.}
\end{center}

\section{Experimental results}
The algorithm performance was tested on the superconducting cavity which was installed in the test cryomodule. The algorithm was implemented on the buncher cavity, which is a normal conducting cavity during the feedforward algorithm development phase.

\subsection{Test result of SRF cavity}
The beam loading effects of the superconducting cavity were measured to compare with a normal conducting cavity. The RF parameters of the SC half-wave resonator are shown in Table~\ref{tab1}.

\begin{center}
\tabcaption{ \label{tab1} RF parameters of the superconducting half-wave resonator cavity.}
\footnotesize
\begin{tabular*}{120mm}{@{\extracolsep{\fill}}ccc}
\toprule RF parameter & Value & Unit \\
\hline
Frequency & 162.5 & MHz \\
$\beta_{opt}$ & 0.1 &  \\
$U_{acc}$ & 0.78 & MV \\
$E_{peak}$ & 25 & MV/m \\
$B_{peak}$ & 50 & mT \\
$Q_0$ & $2.0*10^9$ & \\
$R/Q_0$ & 148 & $\Omega$ \\
$Q_e$(Input coupler) & $1.2*10^6$ & \\
\bottomrule
\end{tabular*}
\end{center}

A dual channel Boonton RF peak power meter~\cite{lab16} was used to measure the forward power of the RF power source and the power of the RF pickup signal from the SC cavity.

\begin{center}
\includegraphics[width=0.7\columnwidth]{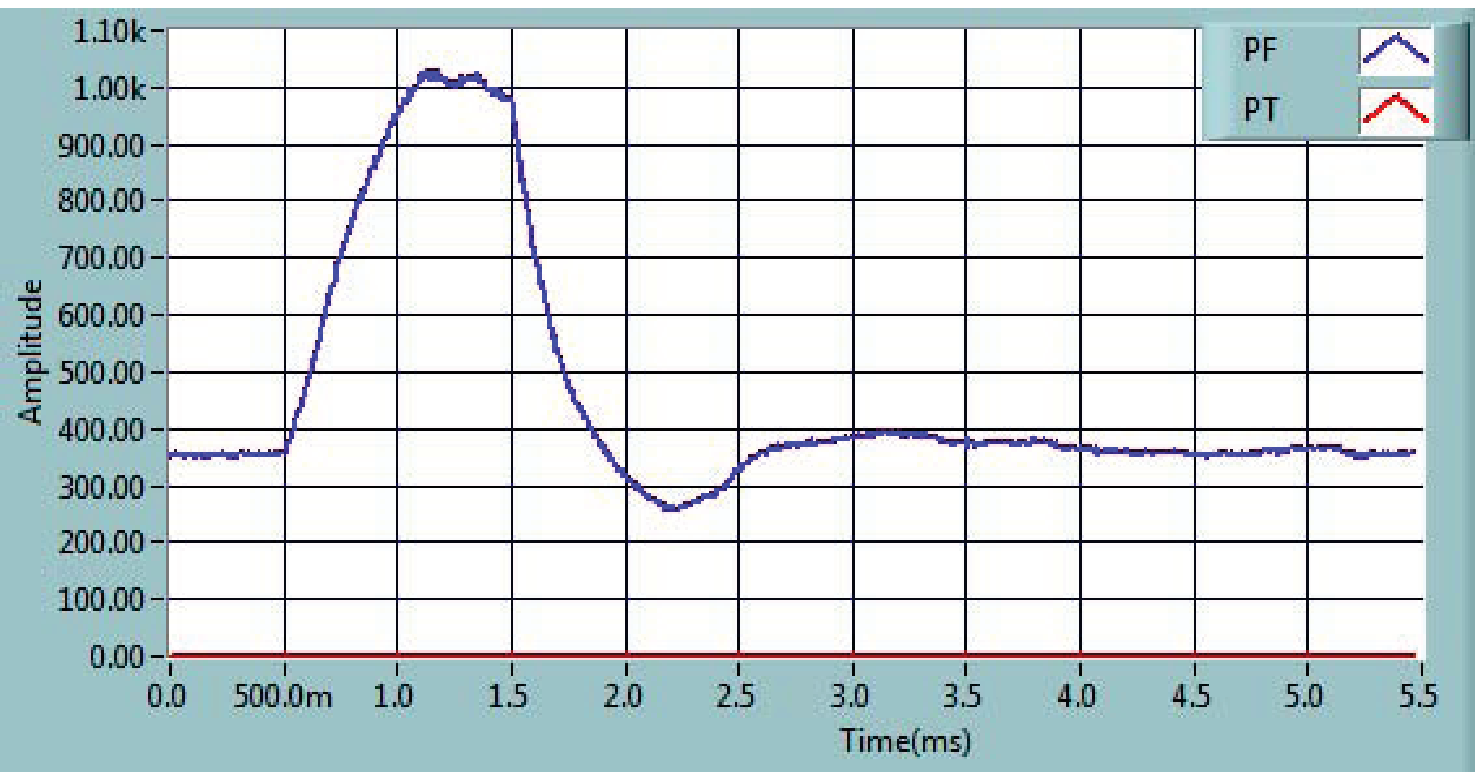}
\figcaption{\label{fig4} Forward RF power waveform of superconducting cavity in closed loop state. The y-axis units are watts.}
\end{center}

\begin{center}
\includegraphics[width=0.7\columnwidth]{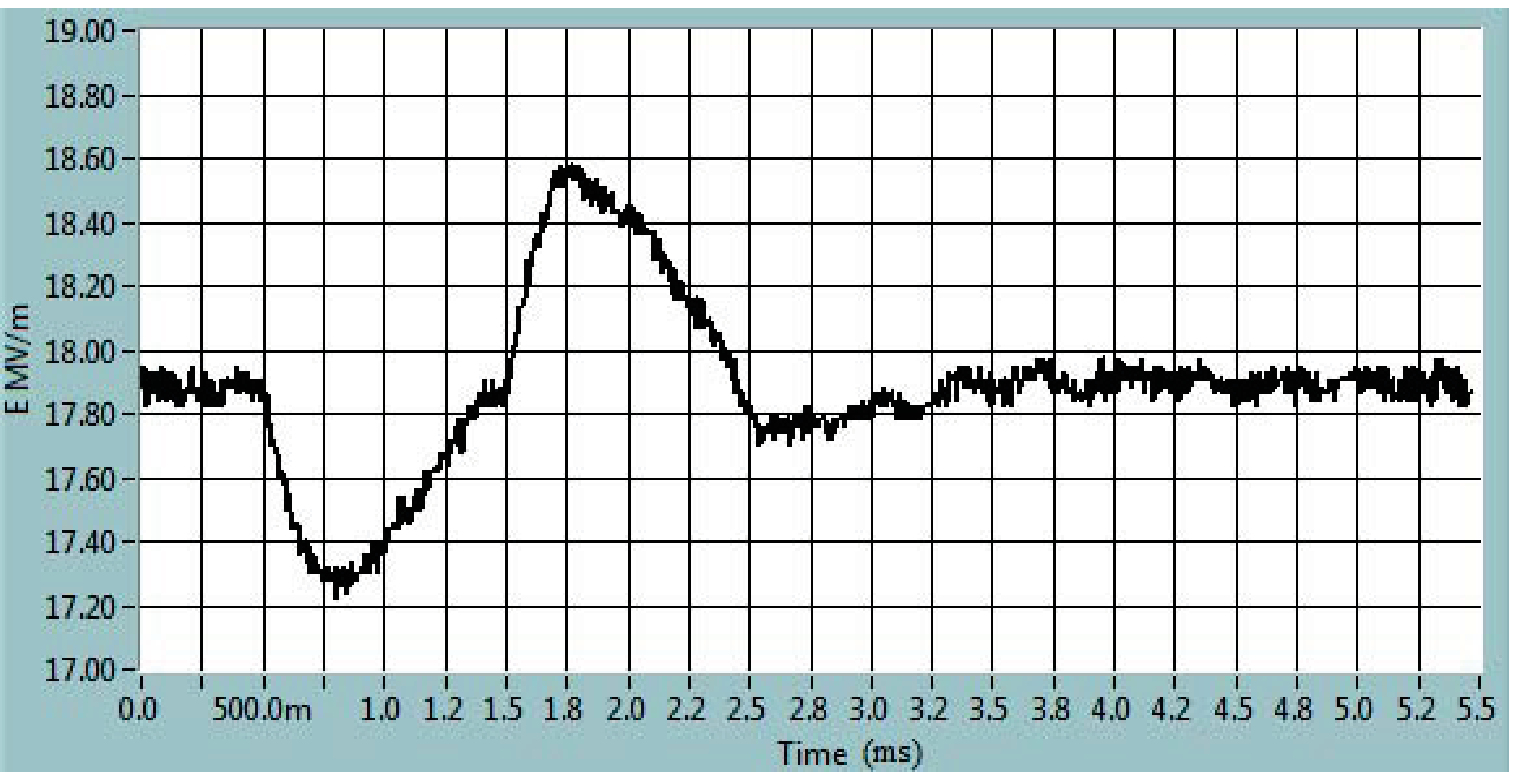}
\figcaption{\label{fig5} Waveform of SC cavity peak electric field.}
\end{center}
Fig.~\ref{fig4} shows the RF forward power waveform of the closed loop state when the FF controller was not enabled. The SC cavity voltage fluctuation was quite large, as can be seen in Fig.~\ref{fig5}. The peak to peak value of cavity electric field variation is shown in Table~\ref{tab2}.

\begin{center}
\tabcaption{ \label{tab2}  LLRF system performance comparison.}
\footnotesize
\begin{tabular*}{120mm}{@{\extracolsep{\fill}}cc}
\toprule LLRF status & Amplitude stability (peak to peak) \\
\hline
No FF control & 7.3\% \\
With FF controller & 1.7\% \\
\bottomrule
\end{tabular*}
\end{center}

As seen in Fig.~\ref{fig7}, the LLRF control system shows a significant improvement in SC cavity field stability after turning on the feedforward controller. The cavity field variation is less than 1.7\%. Figure~\ref{fig6} shows the forward RF power waveform. We are hopeful that this can be improved through implementation of an algorithm which continues to modify the feedforward controller during operation in order to reduce the systematic errors.

\begin{center}
\includegraphics[width=0.7\columnwidth]{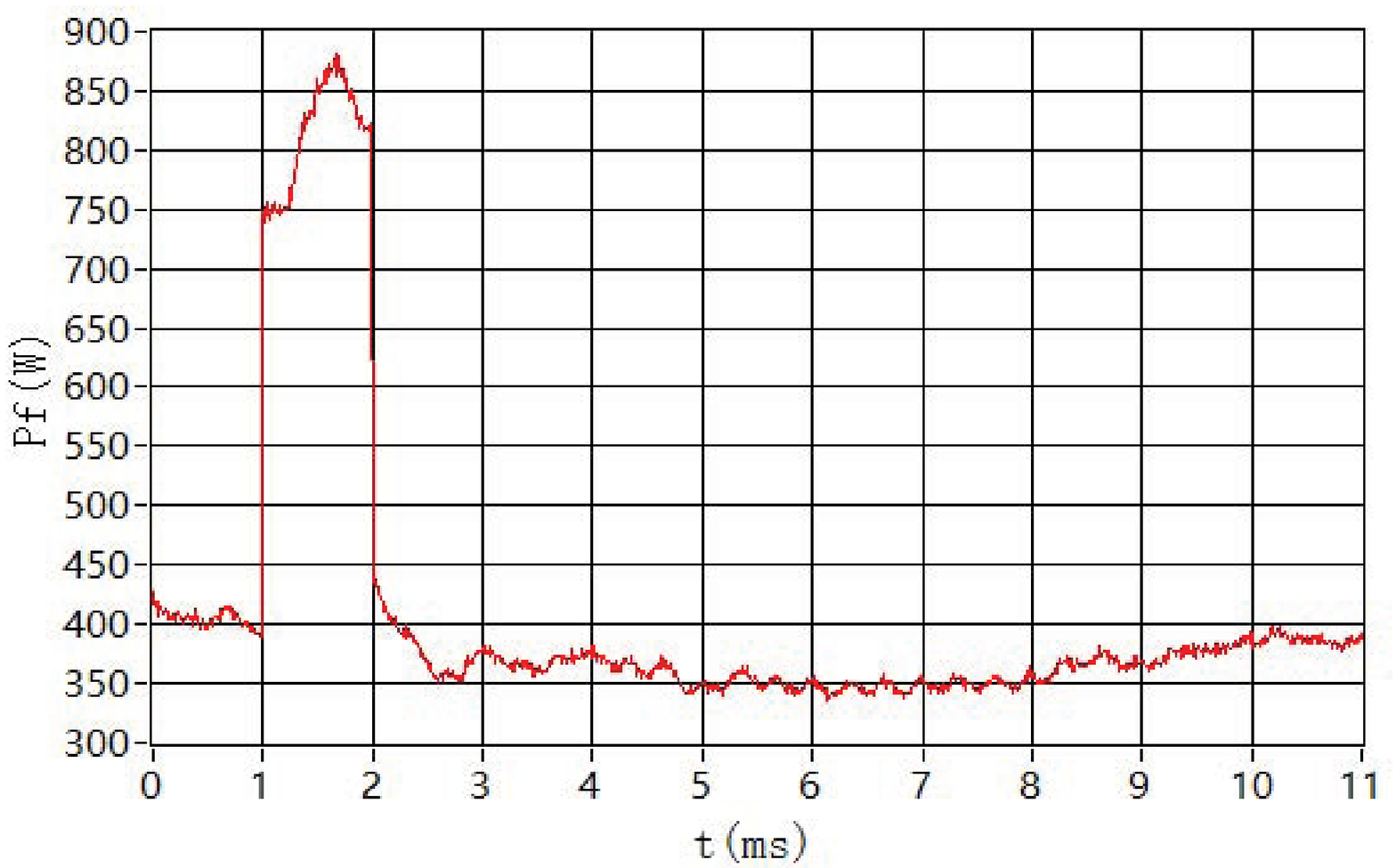}
\figcaption{\label{fig6} Forward power waveform with FF controller running.}
\end{center}

\begin{center}
\includegraphics[width=0.7\columnwidth]{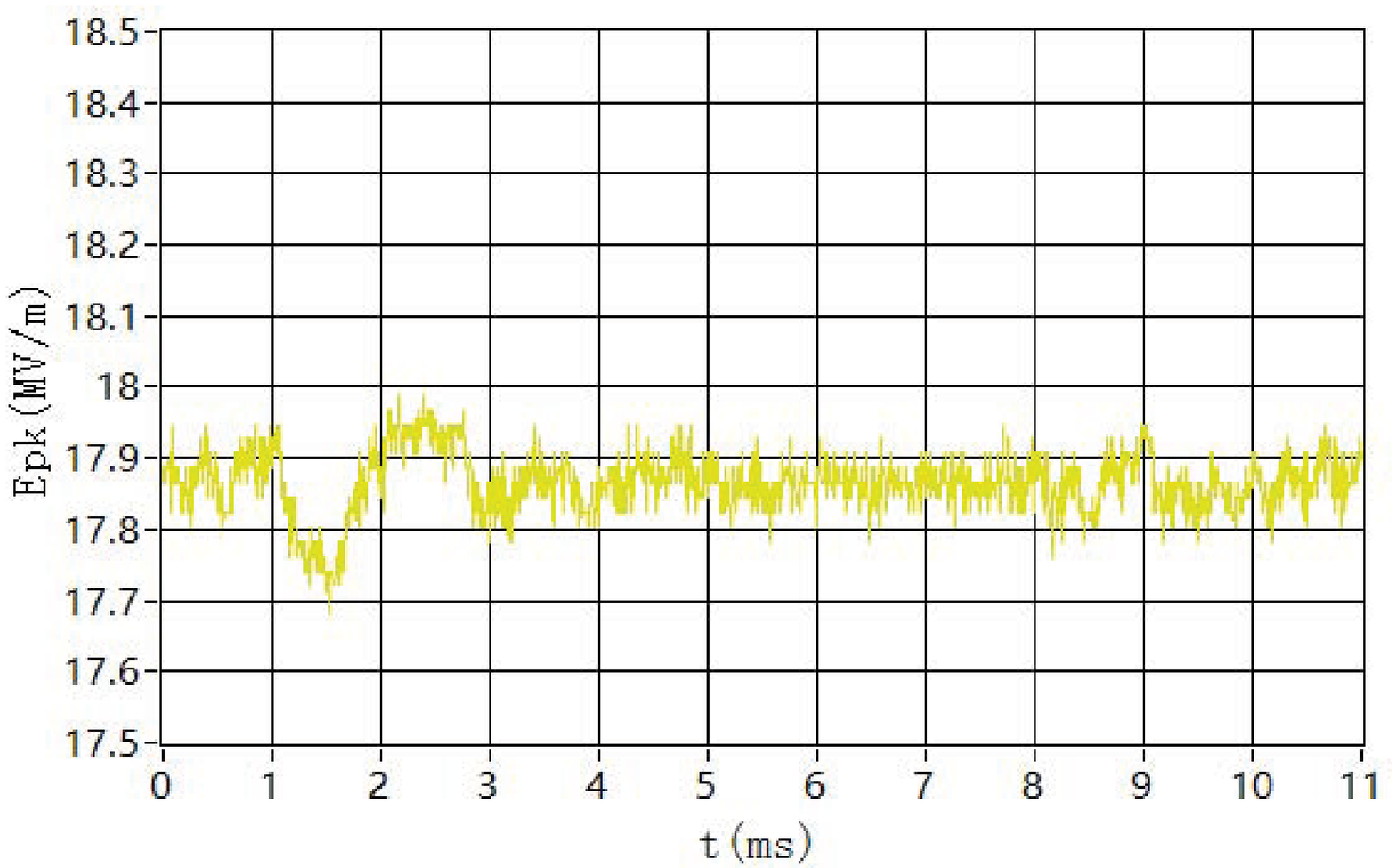}
\figcaption{\label{fig7} SC cavity peak electric field stability improvement with feedforward.}
\end{center}

\subsection{Algorithm verification on buncher}
The beam loading test was first performed on the buncher during development of the RF feedforward control system. The RF parameters of the buncher are shown in Table~\ref{tab3}.

\begin{center}
\tabcaption{ \label{tab3}  RF parameters of the buncher.}
\footnotesize
\begin{tabular*}{120mm}{@{\extracolsep{\fill}}ccc}
\toprule RF parameter & Value & Unit \\
\hline
Frequency & 162.5 & MHz \\
$U_{acc}$ & 100 & kV \\
$\kappa$ & 600 & \\
$Q_0$ & $7733$ & \\
$Q_e$(Input coupler) & $7284$ & \\
\bottomrule
\end{tabular*}
\end{center}

The RF signal from the cavity pickup probe was transferred to an oscilloscope for measurement and analysis. The cavity voltage drop caused by a 2 mA proton beam is shown in Fig.~\ref{fig8}; the yellow waveform is the cavity pickup RF signal, the beam pulse width is 70 $\mu$s, the LLRF control system was in opening loop state, and the buncher was running in accelerating mode. The amplitude variation of the buncher accelerating voltage is about 3.6\%, which is a significant disturbance which can be used to test the feedforward control algorithm.

\begin{center}
\includegraphics[width=0.55\columnwidth]{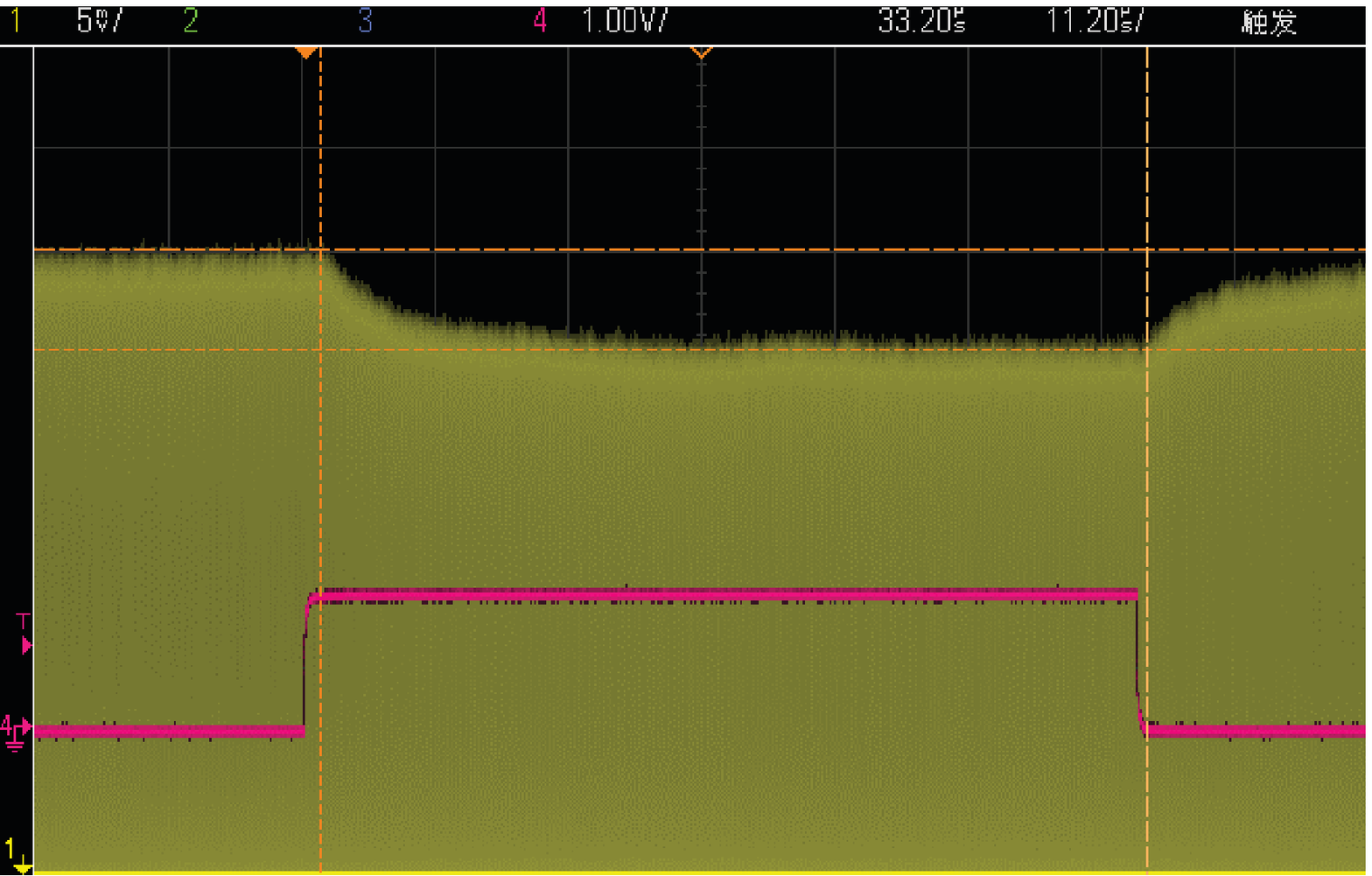}
\figcaption{\label{fig8} Cavity voltage drop of the buncher.}
\end{center}

In order to stabilize the cavity voltage amplitude of the buncher, the LLRF control system operated in closed loop state, but the cavity voltage still fluctuated with beam - there is ¡°droop¡± in the gradient synchronous with the rising edge of the  beam pulse and an ¡°overshoot¡± that begins synchronous with the beam pulse falling edge. This is shown in Fig.~\ref{fig9}. Similar results was observed when using the previous version of the low level RF system.

\begin{center}
\includegraphics[width=0.6\columnwidth]{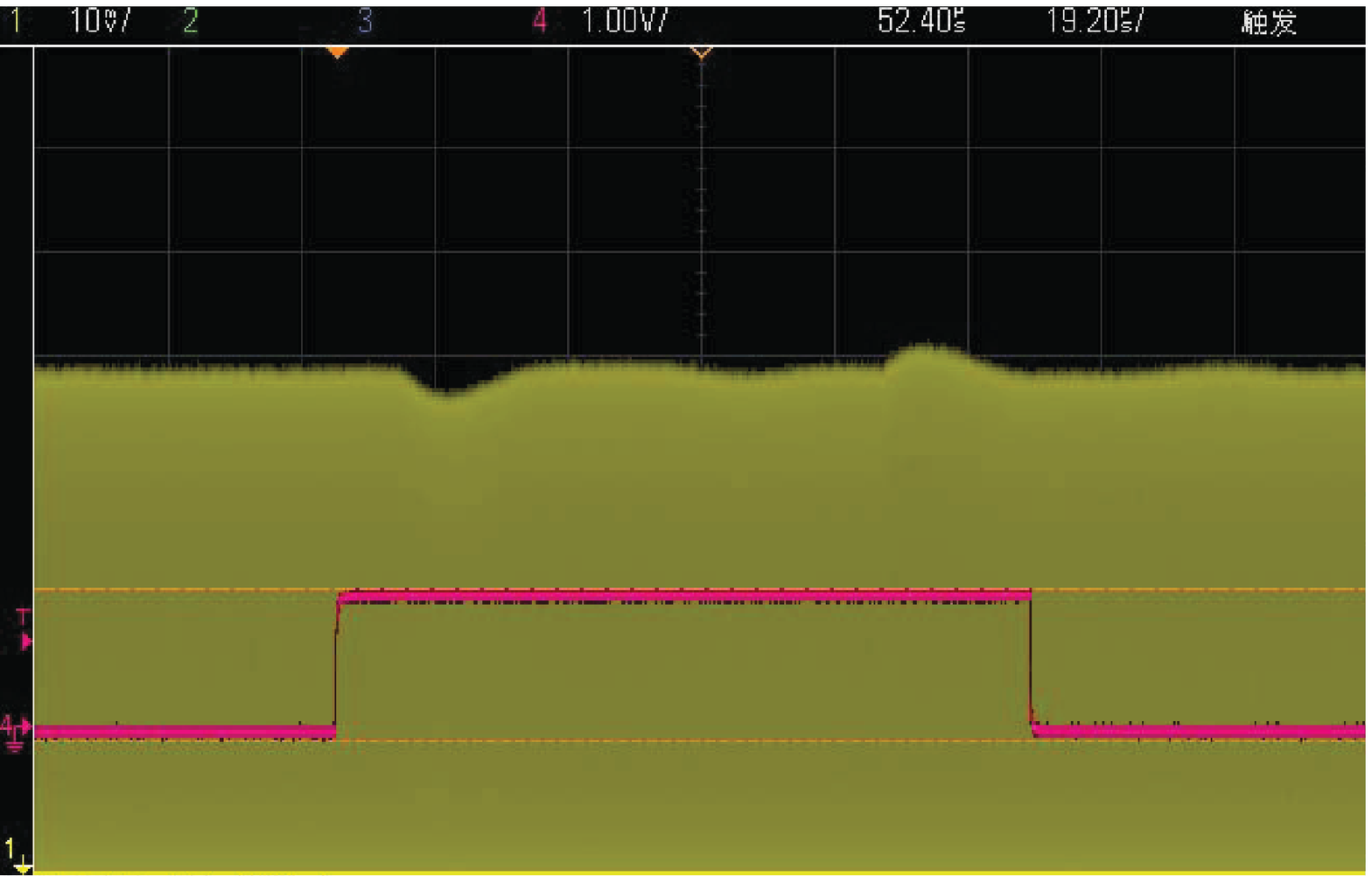}
\figcaption{\label{fig9} Cavity pickup signal of the closed loop state.}
\end{center}

Attempts were made to stabilize the cavity gradient by increasing the PI controller loop gains. However when they were increased above a certain level the system became unstable.

\begin{center}
\includegraphics[width=0.6\columnwidth]{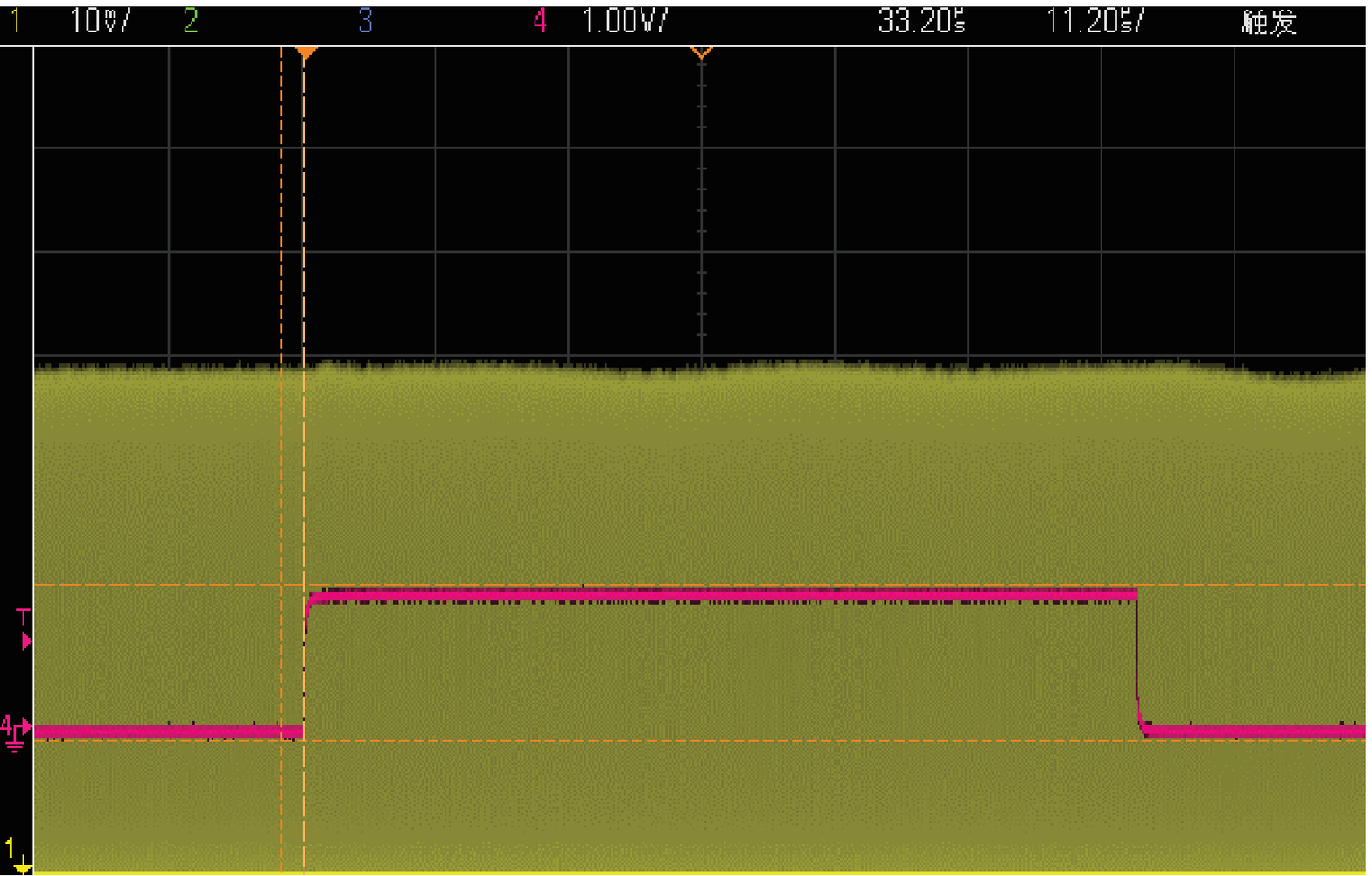}
\figcaption{\label{fig10} Waveform of cavity pickup signal with FF controller running.}
\end{center}

As Fig.~\ref{fig10} shows, there is no obvious repetitive fluctuation after turning on the FF controller. The PI controller was running together with the FF controller in this case, and the cavity pickup signal waveform of this status is shown in Fig.~\ref{fig10}.

\section{Discussion}
These results demonstrate that application of a feedforward control algorothim can be an effective way of compensating transient beam loading. The cavity field stability is substantially improved by applying the algorithm. The feedforward algorithm used in this paper is different from adaptive feedforward control: there is no learning cycle after turning on the feedforward controller. We believe that continuing to apply a learning cycle by averaging over several pulses after the FF control is turned on should further reduce the errors introduced by pulsed beam loading. It should be noted that the trigger of FF control needs to be ahead of the beam pulse because of the cavity response time.

\section{Summary}
A preliminary beam test of the transient beam loading compensation system has been performed, and the repetitive error in the RF control loop caused by the pulsed proton beam was successfully suppressed by using feedback and feedforward control algorithms. Work will continue in this area so as to further improve the algorithm, with the goal of further stabilizing the cavity gradient errors that are driven by pulsed beam loading.

\vspace{-1mm}
\centerline{\rule{100mm}{0.5pt}}
\vspace{2mm}



\clearpage

\end{document}